\documentclass[useAMS,usenatbib]{mn2e}
\usepackage[dvips]{graphicx}
\input{epsf}

\title[Predictions on the high frequency polarization of extragalactic
radio sources]{Predictions on high frequency polarization properties of
extragalactic radio sources and implications for CMB polarization
measurements}

\author[M. Tucci et al.]
  {M.~Tucci,$^1$ E.~Mart\'{\i}nez--Gonz\'alez,$^1$
  L.~Toffolatti,$^2$ J.~Gonz\'alez--Nuevo$^2$ and \\
  \newauthor
  G.~De Zotti$^3$ \\ \\
  $^1$Instituto de F\'{\i}sica de Cantabria, Consejo Superior de
      Investigac\'{\i}ons Cientificas -- Universidad Cantabria,
      Avda. Los Castros s/n, \\ ~39005 Santander, Spain \\
  $^2$Departamento de F\'{\i}sica, Universidad de Oviedo, c/ Calvo
      Sotelo s/n, 33007 Oviedo, Spain \\
  $^3$Osservatorio Astronomico di Padova, INAF, Vicolo
dell'Osservatorio 5, I-35122 Padova, Italy}
\pagerange{\pageref{firstpage}--\pageref{lastpage}} \pubyear{2003}

\def\LaTeX{L\kern-.36em\raise.3ex\hbox{a}\kern-.15em
    T\kern-.1667em\lower.7ex\hbox{E}\kern-.125emX}

\begin{document}

\label{firstpage}

\maketitle

\begin{abstract}
We present a method to simulate the polarization properties of
extragalactic radio sources at microwave frequencies. Polarization
measurements of nearly $2\times10^6$ sources at 1.4\,GHz are
provided by the NVSS survey. Using this catalogue and the GB6
survey, we study the distribution of the polarization degree of
both steep-- and flat--spectrum sources. We find that the
polarization degree is anti--correlated with the flux density for
the former population, while no correlation is detected for the
latter. The available high--frequency data are exploited to
determine the frequency dependence of the distribution of
polarization degrees. Using such information and the evolutionary
model by \citet{tof98}, we estimate the polarization
power spectrum of extragalactic radio sources at $\geq 30\,$GHz
and their contamination of CMB polarization maps. Two distinct
methods to compute point--source polarization spectra are
presented, extending and improving the one generally used in
previous analyses. While extragalactic radio sources can
significantly contaminate the CMB E-mode power spectrum only at
low frequencies ($\nu\la30$\,GHz), they can severely constrain the
detectability of the CMB B--mode up to $\nu\simeq100\,$GHz.
\end{abstract}


\section{Introduction}
\label{s1}

In the last year, two experiments provided the first observational
evidence of the polarization of the Cosmic Microwave Background
(CMB) radiation: DASI \citep{kov02} achieved a direct measure of
E--mode polarization (see \citealt{kam97} and \citealt{zal97} for
a definition of E and B modes), while WMAP \citep{ben03a} detected
the cross--correlation between the CMB temperature anisotropies
and E--mode polarization, giving an estimate of the reionization
optical depth \citep{kog03}. A large number of experiments have
been planned to measure the CMB polarization (see \citealt{cec02};
\citealt{han03}), motivated by its huge information content.
Besides probing the ionization history, the CMB polarization can
open a window on the primordial phase of the Universe through the
detection of the B--mode, i.e. the curl component of the
polarization field. In fact, the B--mode polarization is generated
by tensor metric perturbations and, in inflation models, its
amplitude is directly proportional to the energy scale at which
the inflation has occurred. However, the detection of this
component is a real challenge, not only due to the very low level
of the signal, but also to the presence of foregrounds and effects
that can mix the E-- and B--modes of the CMB polarization. For
example, the gravitational lensing produced by large scale
structures converts a fraction of the CMB E--mode component to the
B--mode one. Although the effect is of a few percent or below, the
lensing--induced curl mode can overwhelm the primordial one
(\citealt{zal98}; \citealt{kno02}).

Our knowledge of foreground polarization is still poor. Among the
Galactic emissions which dominate foreground intensity
fluctuations on large angular scales, free-free is nearly
unpolarized while synchrotron is highly polarized; its polarization
properties have been studied at GHz frequencies (\citealt{tuc00},
2002; \citealt{bac01}; \citealt{bru02}), but we
have little or no direct information at frequencies of interest
for CMB studies. At mm wavelengths, the Galactic polarized signal
is due essentially to the dust emission. Estimates of this
contribution have been so far provided only by models based on HI
maps \citep{pru98} or on the starlight polarization \citep{fos02},
although direct sub-mm measurements over a substantial fraction of
the sky have been recently provided by the ARCHEOPS experiment
\citep{ben03}.

In this paper we deal with extragalactic radio sources, which are
expected to be the dominant polarized foreground on small angular
scales. Accurate studies of their contribution to intensity
fluctuations, based on evolutionary models fitting the available
data at many frequencies, have been published. In our analysis we
adopt the \citet{tof98} (hereafter T98) model in order to predict
the number counts of extragalactic radio sources at cm and mm
wavelengths. This model reproduces very well the number counts of
different classes of bright radio sources at GHz frequencies (see
Sections \ref{s2} and \ref{s3}); moreover, the accuracy of its
high--frequencies predictions has been confirmed by the recent VSA
(\citealt{tay01}; \citealt{wal03}), CBI \citep{mas02} and WMAP
\citep{ben03b} surveys between 15 and 44\,GHz \citep{tof03}. More
specifically, the WMAP survey has detected 208 sources in a sky
area of 10.38\,sr at $\vert b^{II}\vert
>10\degr$ at flux densities $S\ge0.9$--1.0\,Jy. The T98 model predicts
270--280 sources in the same sky area, so that the average offset
is a factor of $\sim 0.75$. Moreover the distribution of spectral
indexes in the WMAP sample peaks around $\alpha=0.0$, which is
exactly the mean spectral index for flat--spectrum sources adopted
by T98, and the fraction of steep--spectrum (i.e., $\alpha> 0.5$,
$S\propto \nu^{-\alpha}$) sources is of $\sim 12\%$, to be
compared with a predicted fraction of $\simeq 10\%$. As a
curiosity, we also note that the brightest source detected by WMAP
has a flux density of $S\simeq 25$\,Jy which corresponds exactly
to the value for which the model predicts 1 source all over the
sky.

Most of the available information on polarization properties of
extragalactic sources refers to GHz frequencies, while for CMB
studies we are interested in frequencies $\ga 30\,$GHz. There are
at least two reasons to expect polarization properties to be
frequency dependent, and, in particular, higher polarization at
higher frequencies. First, at low frequencies substantial
depolarization may be induced by Faraday rotation. Determinations
of rotation measures for radio galaxies and quasars yield values
from tens to $10^3$\,rad\,m$^{-2}$ (\citealt{ode89};
\citealt{tay00}; \citealt{pen00}; \citealt{mes02}, henceforth
M02), implying that depolarization may be significant up to
$\nu\simeq 10$\,GHz. Second, especially in compact objects, as the
observing frequencies increase emitting regions are
closer and closer to the nucleus, where the degree of the ordering
of magnetic fields and, as a consequence, the polarization degree,
may be higher and higher.

\begin{figure}
\includegraphics[width=84mm]{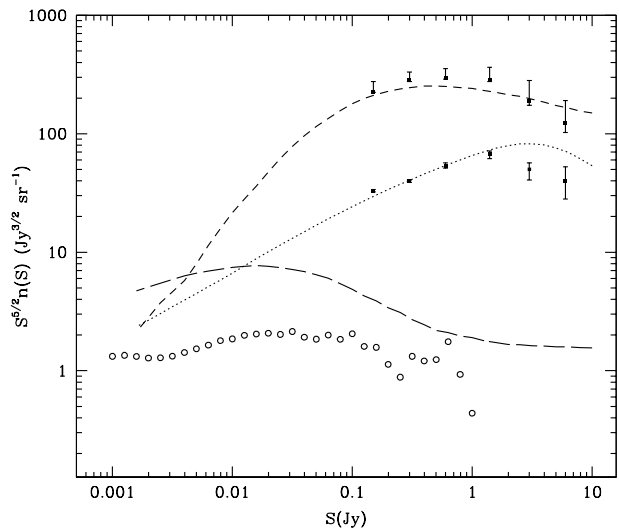}
\caption{Differential number counts, normalized to $S^{5/2}$, of
steep- (upper data points) and flat--spectrum (lower data points)
NVSS sources, compared with predictions of the T98 model
(steep--spectrum sources: short dashed curve; flat--spectrum
sources: dotted curve; ``starburst'' galaxies: long dashed curve).
The upper part of error bars for steep spectrum sources allows for
the possibility that NVSS sources missing a GB6 counterpart belong
to this population. The dots show the total differential counts in
terms of polarized flux. } \label{f1}
\end{figure}

Preliminary estimates of the contribution of extragalactic radio
sources to the polarization angular power spectrum (APS) were
worked out by \citet{dez99} and M02. \citet{dez99} simply scaled
the temperature power spectrum by the mean squared polarization
degree of radio sources, assuming a Poisson spatial distribution.
M02, instead, computed the APS directly from the NVSS polarimetric
data \citep{con98}, estimating a factor of 3 increase in the
polarization degree of steep--spectrum sources from 1.4 GHz to
$\geq 20\,$GHz and assuming that, for flat--spectrum sources, the
polarization degree either is frequency independent or
increases by a factor of 3.

In this paper, we improve on previous estimates by studying the
probability distribution of the polarization degree for both
steep-- and flat--spectrum sources at $\nu\le15\,$GHz, providing a
recipe to extrapolate the distributions to higher frequencies. Our
results consist of a technique to generate simulated catalogues of
extragalactic radio sources and also in accurate expressions to
estimate the polarization APS, after having removed the strongest
polarized sources.

\section{Polarization properties of NVSS sources}
\label{s2}

The starting point of the analysis is the NRAO VLA Sky Survey
(NVSS; \citealt{con98}). This survey covers $\Omega\simeq10.3\,$sr
of the sky with $\delta\ge-40\degr$ at 1.4 GHz, providing the flux
density $S$ and the Stokes parameters $Q$ and $U$ of almost
$2\times10^6$ discrete sources with $S\ga2.5\,$mJy. The images
obtained from interferometric data have $\theta=45\arcsec\,$FWHM
resolution and the rms brightness fluctuation in the $Q$ and $U$
parameters is $\sigma\simeq0.29\,$mJy\,beam$^{-1}$.

\begin{table*}
\begin{minipage}{175mm}
\caption{Polarization degree at $1.4\,$GHz of steep-- and
flat--spectrum sources.} \label{t2}
\begin{tabular}{@{}cccccccccccc}
\hline
     & & & flat & & & & & steep & & \\
flux(mJy) & N & N(T98) & N$_{\Pi<1\%}(\%)$ & median($\%$) &
mean($\%$) & N & N(T98) & N$_{\Pi<1\%}(\%)$ & median($\%$) &
mean($\%$) \\
\hline
100--200 & 2305 & 2144 & 43.3 & 1.33 & 2.16 & 15621 & 15409 &
35.8 & 1.77 & 2.70 \\
200--400 &  980 & 1045 & 43.5 & 1.24 & 2.01 & 6925 & 6006 & 40.0
& 1.52 & 2.40 \\
400--800 &  473 & 486  & 40.2 & 1.50 & 1.95 & 2552 & 2428 & 41.2
& 1.44 & 2.25 \\
 $>800$  &  260 & 364  & 45.7 & 1.12 & 1.92 & 1096 & 1165 & 47.0
& 1.14 & 2.02 \\
\hline
\end{tabular}
\end{minipage}
\end{table*}

Information on the spectral index $\alpha$ of NVSS sources is
obtained exploiting the Green Bank 4.85\,GHz survey (GB6,
\citealt{gre96}), which covers the declinations
$0\degr<\delta<75\degr$ ($\Omega\simeq6.07\,$sr) to a flux limit
of $S_{4.85}=18\,$mJy with a resolution FWHM$=3.5$\,arcmin. We
have cross--matched the positions of NVSS sources with
$S_{1.4}\ge100\,$mJy with the GB6 catalogue, taking all matches
with position offsets less than 3 times the uncertainty on the GB6
position. Sources with Galactic latitude $|b|\le2\degr$ have been
rejected, in order to guarantee that nearly all the sources are
extragalactic. The $100\,$mJy flux limit has been chosen to have
$\ga3\sigma$ detections of polarization down to a percent level.
About $86\%$ of NVSS sources turned out to have a GB6
counterpart. Whenever more than one NVSS source falls within the
GB6 beam, as a consequence of the better NVSS angular resolution
which may lead to individually resolve multiple components, we
have summed up their fluxes, corrected for the effect of the GB6
beam. We ended up with a sub--sample of 29299 NVSS sources, of
which $\sim87\%$ are steep--spectrum ($\alpha>0.5$) and $\sim13\%$
are flat--spectrum ($\alpha\le0.5$).

In Fig. \ref{f1} we show the differential number counts of NVSS
sources as a function of the total and linearly polarized
($PI=\sqrt{Q^2+U^2}$) flux density. The predictions of the
T98 model (for a flat CDM cosmology with $\Omega_{\Lambda}=0.7$
and $h=0.7$) are also displayed. According to the model, the
dominant contribution at high and moderate fluxes comes from
steep--spectrum sources, while ``starburst'' galaxies are
negligible for $S$ above a few tens of mJy. The fraction of
flat--spectrum sources increases with flux density and is important
at the Jy level. As shown by Table \ref{t2} and by Fig. \ref{f1}, the
observed number of flat-- and steep--spectrum sources in total
flux--density bins compares favourably with the predictions, N(T98),
of the T98 model. The latter only weakly overestimates the observed
number of flat--spectrum sources for $S>800\,$mJy; such discrepancy
will be taken into account when computing their contribution
to the polarization power spectrum in Section \ref{s5}.

It is interesting to note that the normalized $PI$ number counts
keep nearly constant over all the flux density range where they
are defined. The different shape compared with the total power number
counts indicates a dependence of the polarization degree on flux
density (see below).

\begin{figure}
\includegraphics[width=84mm]{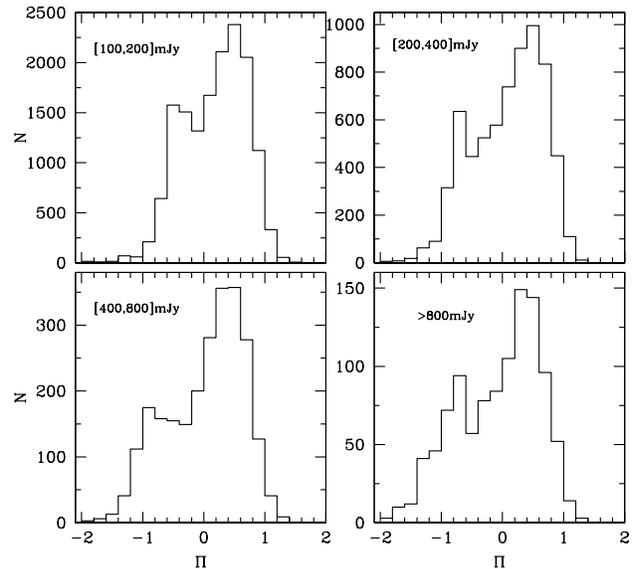}
\caption{Polarization degree distribution of steep--spectrum NVSS
sources for the flux density intervals specified in each panel.}
\label{f2}
\end{figure}

\begin{figure}
\includegraphics[width=84mm]{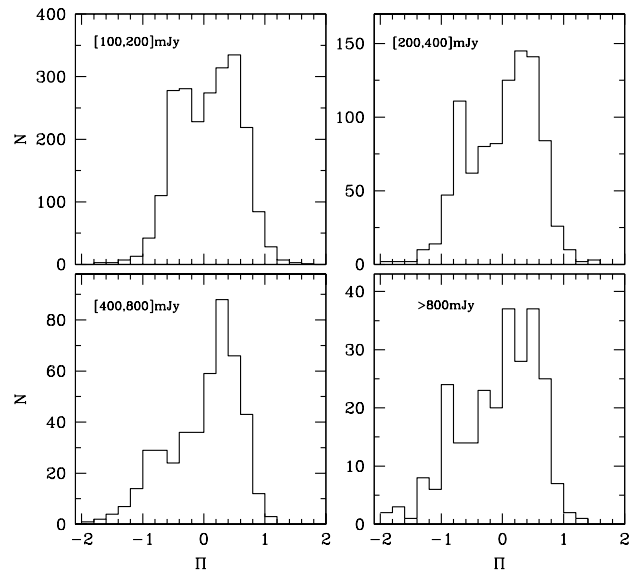}
\caption{As in Figure \ref{f2} but for flat--spectrum sources.}
\label{f3}
\end{figure}

Following the analysis of M02, we study the distribution of the
percentage polarization degree ($\Pi=100PI/S$) at 1.4 GHz for
steep-- and flat--spectrum sources for different flux--density
intervals (see Figs. \ref{f2} and \ref{f3}). Table \ref{t2}
reports the median and the mean value of $\Pi$ for such
distributions and the percentage of sources with $\Pi$ lower than
$1\%$ (N$_{\Pi<1\%}$). The low--$\Pi$ tail of distributions is
contaminated by noise and residual instrumental polarization;
however, while the relevance of the noise decreases when the
polarized intensity increases, the latter is proportional to the
flux density of sources. \citet{con98} found that the instrumental
polarization is $\approx0.12\%$ of $S$ for a large sample of
sources stronger than 1\,Jy and estimated that, in any case, it
should be less than $0.3\%$.

The results of Table \ref{t2} highlight an anti--correlation
between the polarization degree and the flux density for
steep--spectrum sources: the median value of $\Pi$ steadily
decreases from $1.8\%$ at 100--200\,mJy to $1.1\%$ at
$S\ge800\,$mJy (we consider the median because it is less affected
than the mean by instrumental polarization). A similar trend is
not found for flat--spectrum sources, whose median values show
only small fluctuations, all compatible with a constant value at
the 1-$\sigma$ level. Pearson's linear correlation coefficient $r$
between the flux density and the polarization degree of sources is
low for both flat (-0.017) and steep--spectrum (-0.034) sources,
yielding a $30\%$ probability of the null hypothesis (i.e., no
correlation) in the flat case. However, the $\chi^2$ test clearly
rules out the possibility that the polarization degree
distributions of steep--spectrum sources in the range
100--200\,mJy and at higher flux density are drawn from the same
parent distribution (probabilities of the order or less than
$10^{-4}$; see Table \ref{t3}). On the contrary, in the case of
flat--spectrum sources, the test indicates consistency with the
same parent distribution for all flux density ranges (in
particular, a probability of $95\%$ is found comparing the
intervals at 100--200\,mJy and $>800$\,mJy).

Differences in the $\Pi$ distributions for flat--spectrum sources
can be perceived in Figure \ref{f3} for $\Pi<1\%$. However, they
are probably not intrinsic but induced by instrumental effects: at
low fluxes (100--400\,mJy) only few sources are detected with
$\Pi\la0.1\%$ because the noise rms contribution is $\simeq
0.3\%$; viceversa, at $S>400$\,mJy, where the noise is practically
negligible, a significant tail of very low values of $\Pi$ is
observed. Finally, we note that artificial peaks in the
distributions at $\Pi\simeq0.1$--$0.25\%$ are produced by the
residual instrumental polarization.

\begin{table}
\begin{minipage}{80mm}
\caption{Results of the $\chi^2$ test for $\Pi$ distributions at
different fluxes. The distributions are binned in intervals of the
$1\%$ width. Probability values above the diagonal refer to
flat--spectrum sources, those below the diagonal, to
steep--spectrum sources.} \label{t3}
\begin{tabular}{@{}ccccc}
\hline
 (mJy)   & 100--200 & 200--400 & 400--800 & $>800$ \\
100--200 &    --    & 0.28     & 0.91     & 0.95   \\
200--400 & $2.1\times10^{-4}$ &    --    & 0.12     & 0.06   \\
400--800 & $1.8\times10^{-4}$ & $4.8\times10^{-2}$ & -- & 0.52   \\
 $>800$  & $1.4\times10^{-6}$ & $6.1\times10^{-3}$ & 0.16 & -- \\
\hline
\end{tabular}
\end{minipage}
\end{table}

\begin{figure}
\includegraphics[width=84mm]{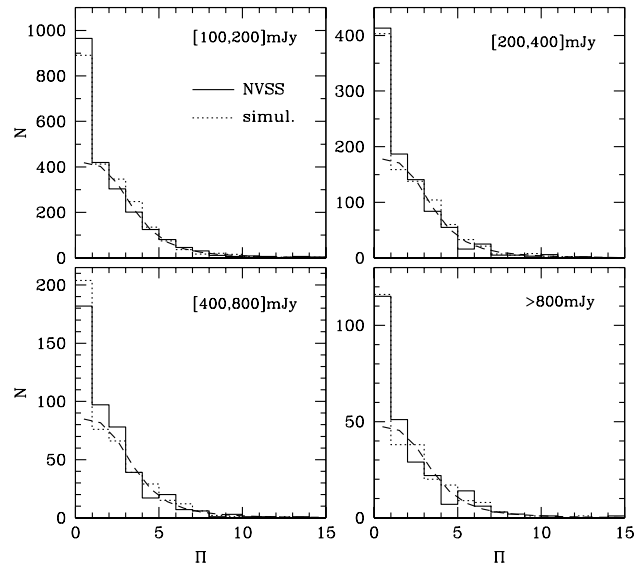}
\caption{The $\Pi$ distribution for NVSS flat--spectrum sources (solid
histogram) compared with results of the simulated catalogue (dotted
histogram). The dashed curves are the fit given by the first term of
Eq. (\ref{e1}).}
\label{f4}
\end{figure}

\begin{figure}
\includegraphics[width=84mm]{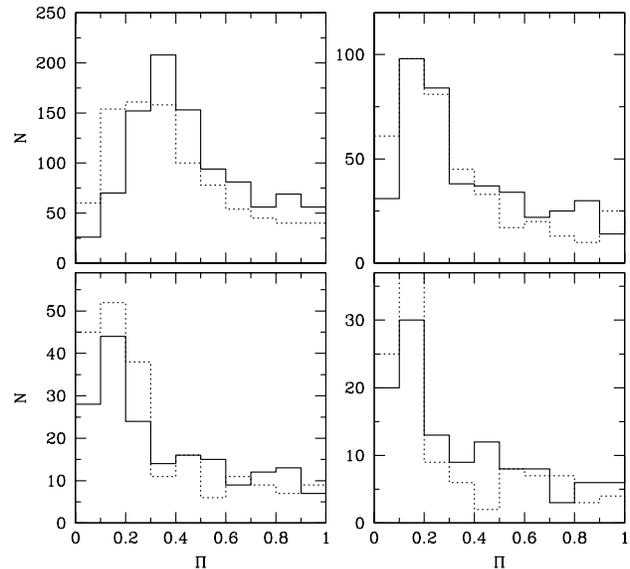}
\caption{The same as in Figure \ref{f4} but for $\Pi=[0,1]\%$.}
\label{f5}
\end{figure}

The origin of the anti--correlation in steep--spectrum sources is
not clear yet. A possible explanation is a systematic increase of
the mean redshift of sources with decreasing the flux density,
which would entail a decrease of the Faraday rotation measure
$\propto (1+z)^{-2}$. A moderate anti-correlation between mean
redshift and flux density is indeed expected based on the T98
model. On the other hand, currently available data do not provide
any evidence of a positive correlation between polarization degree
and redshift of sources.

\begin{table*}
\begin{minipage}{150mm}
\caption{Polarimetric data on extragalactic radio sources.}
\label{t1}
\begin{tabular}{@{}lccl}
\hline
Reference & $\nu$(GHz) & N. sources & Comments \\
\hline
\citet{eic79} & [0.4,\,15]    & 510  & multi--frequency data\\
\citet{tab80} & [0.4,\,10.7]  & 1510 & multi--frequency data\\
\citet{sim81} & [1.6,\,10.5]  & 555  & multi--frequency data\\
\citet{per82} & 1.5,\,4.9     & 404  & \\
\citet{rud85} & [1.4,\,90]    & 20   & flat; multi--frequency data\\
\citet{all92} & 4.8,\,14.5    & 62   & 35 flat, 27 steep; complete
sample ($S_{5{\rmn GHz}}>1.3$\,Jy)\\
\citet{oku93} & 10            & 99   & flat; complete sample
($S_{5{\rmn GHz}}>0.8$\,Jy)\\
\citet{nar98} & 273           & 26   & flat\\
\citet{con98} & 1.4           & $\sim2\times10^{6}$ & complete sample
($S_{1.4{\rmn GHz}}>2.5$\,mJy)\\
\citet{all99} & 4.8,\,14.5    & 41   & BL\,LAC\\
\citet{zuk99} & 4.75          & 154  & 28 flat, 122 steep\\
\citet{lis01} & 43            & 30   & flat; 90$\%$ complete sample
($S_{5{\rmn GHz}}>1.3$\,Jy)\\
VLA Calibrators & 5,\,8.5,\,22,\,43 & 62   & 55 flat, 7 steep\\
\hline
\end{tabular}
\end{minipage}
\end{table*}

Contrary to M02, we do not find any evidence that the
polarization of flat--spectrum sources depends on the flux
density. The discrepancy is probably a consequence of the
different separation in steep-- and flat--spectrum sources. In M02,
the fraction of NVSS flat--spectrum sources is very high
($\sim44\%$), and roughly constant at every flux--density
interval. It is possible that, not having taken into account the
effect of the different angular resolution in the comparison
between NVSS and GB6 flux densities, M02 have underestimated the
spectral index of a significant fraction of steep--spectrum
sources, misclassifying them as flat--spectrum.

\subsection{The 1.4\,GHz $\Pi$ probability distribution for
flat--spectrum sources}

The Stokes parameters $Q$ and $U$, measured by the NVSS, are the
sum of three terms:
\begin{equation}
Q,U(\rmn{obs})=Q,U(\rmn{int})+Q,U(\rmn{noise})+Q,U(\rmn{res})\,,
\label{e0}
\end{equation}
where $Q,U(\rmn{int})$ are the intrinsic values, while the other
two terms represent the contribution of the noise and of the
residual instrumental polarization. We find that the distribution
of $Q,U(\rmn{obs})$ for flat--spectrum sources can be reproduced
adopting the following expression for the intrinsic distribution
of polarization degrees at 1.4\,GHz $\Pi_{1.4}$ (considering the flux
density from the NVSS catalogue and a random distribution of
polarization angles):
\begin{equation}
{\mathcal P}(\Pi_{1.4})={a \over
2.7+0.025\Pi_{1.4}^{3.7}}+bf(\Pi_{1.4})\,,
\label{e1}
\end{equation}
with
\begin{displaymath}
f(\Pi_{1.4})=\left\{\begin{array}{ll}
{1 \over 0.1} & \textrm{if $\Pi_{1.4}\le0.1$}\\
0 & \textrm{if $\Pi_{1.4}>0.1$}
\end{array} \right.
\label{e1a}
\end{displaymath}
The values of the coefficients  ($a=0.51$,  $b=0.24$) are
determined by the condition that the observed fraction of sources
with $\Pi_{1.4}<1\%$ has to be reproduced. The first term in
Eq.~(\ref{e1}) is the best fit for the $\Pi_{1.4}$ distribution of
sources with $100\le S<200$\,mJy and $\Pi_{1.4}\ge1\%$ (it is shown by
the dashed lines in Figure \ref{f4}). The second term corresponds
to a population of nearly unpolarized sources. In fact, for about
$20\%$ of the sources, the polarization degree given by the NVSS
catalogue is less than $0.4\%$: these values are only upper limits
to the intrinsic polarization since they can be accounted for by
instrumental polarization and noise. For such sources we have
assumed an intrinsic polarization uniformly distributed between 0
and 0.1$\%$.

The distributions of the noise and of the residual instrumental
polarization for $Q$ and $U$ are assumed to be Gaussian with zero mean
and variance 0.29\,mJy and $0.001S$, respectively.

Figures \ref{f4} and \ref{f5} compare the polarization degree
distributions of NVSS sources with the results of our simulations.
The good agreement at every flux--density interval is evident and
it is confirmed by the Kolmogoroff--Smirnov (K--S) test. The
probability that the two sets of data come from the same
distribution is: $6\times10^{-3}$, 0.23, 0.30, 0.60 for
100--200\,mJy, 200--400\,mJy, 400--800\,mJy, $>800$\,mJy
respectively. The fit is not satisfactory only for the low
polarization portion ($\Pi\la0.5\%$) of the distribution for the
100--200\,mJy interval (see Figure \ref{f5}), probably due to our
difficulty in correctly modelling the instrumental polarization.

\section{Polarization data for Extra-galactic Radio Sources at
$\nu>1.4$\,GHz}
\label{s3}

Data on the polarization of extragalactic radio sources at
frequencies higher than 1.4\,GHz are rather limited, particularly
above 5 GHz. At high frequencies, where Faraday depolarization
should be negligible, data are restricted to incomplete samples of
few tens of objects (see Table \ref{t1}).

Complete samples are essential to avoid selection biases. For this
reason we have extracted from the MIT--Green Bank surveys
(\citealt{ben86}, \citealt{lan90}, \citealt{gri90},
\citealt{gri91}) a complete sub--sample of sources with $S_{5{\rmn
GHz}}>1.4\,$Jy in the following sky areas:
\[
|b|>10\degr\,,~~~~-0\degr.5<\delta<19\degr.5
\]
\[
5^h\le {\rmn {ra}}<19^h\,,~~20^h<{\rmn
{ra}}<4^h\,,~~~~17\degr<\delta<39\degr.15
\]
\[
15^h.5<{\rmn {ra}}<19^h\,,~~20^h<{\rmn
{ra}}<2^h.5\,,~~~~37\degr<\delta<51\degr\,,
\]
and we complemented it with sources from the Pearson--Readhead
survey (\citealt{pea81}, 1988) at $\delta>35\degr$ and
$|b|>10\degr$. Polarization measurements at 4.8\,GHz were found
for 139 ($\sim95\%$) of sources in the final sample (hereafter MG+
sample). Each source was then identified in the NVSS and the
spectral index between 1.4 and 4.8\,GHz was estimated.

\begin{figure}
\includegraphics[width=84mm]{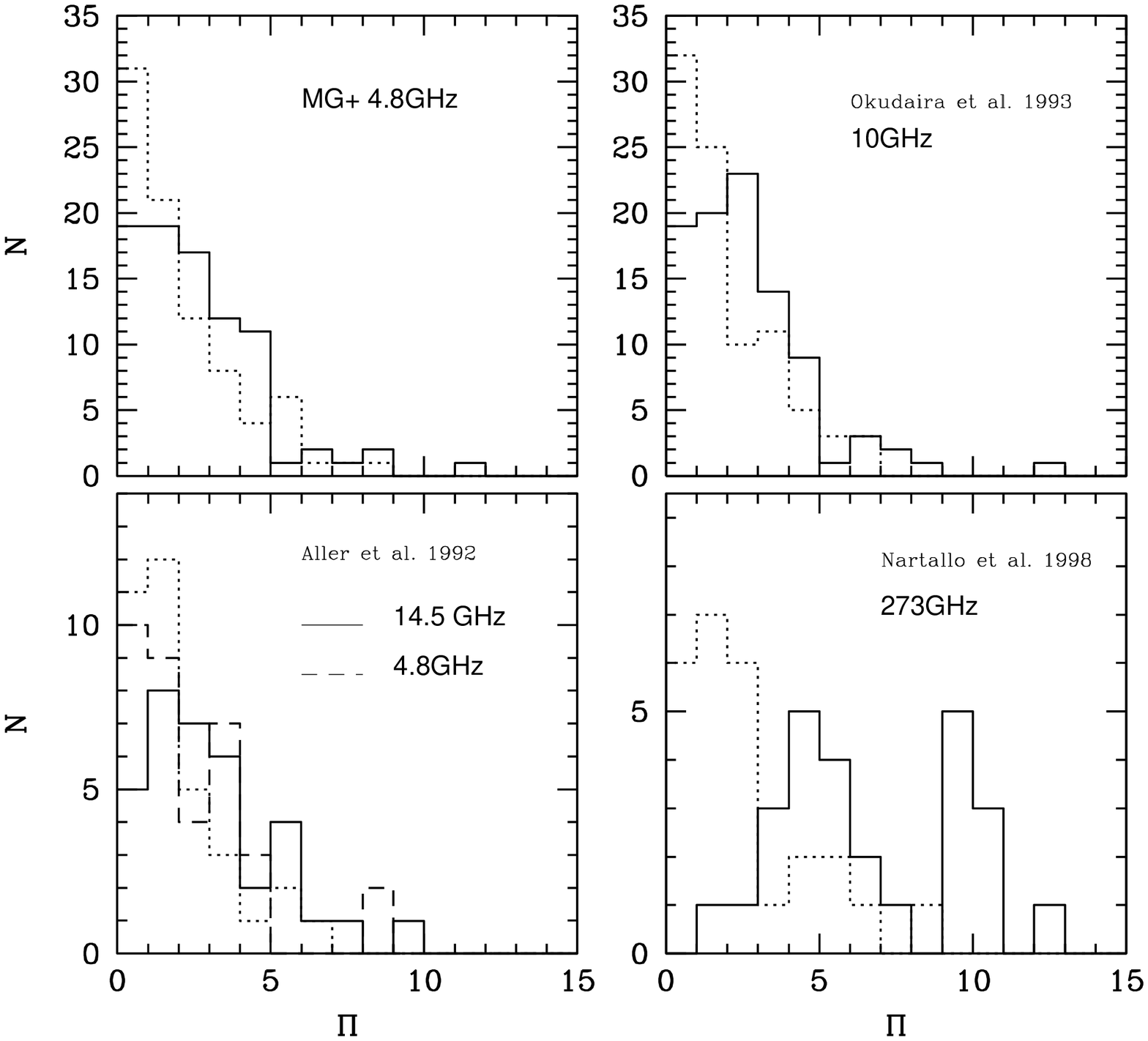}
\caption{$\Pi$ distribution of flat--spectrum sources for
different samples at $\nu>1.4\,$GHz (solid histograms). Dotted
histograms show, for comparison, the $\Pi$ distributions of the
same  sources at 1.4\,GHz, based on NVSS measurements.} \label{f6}
\includegraphics[width=84mm]{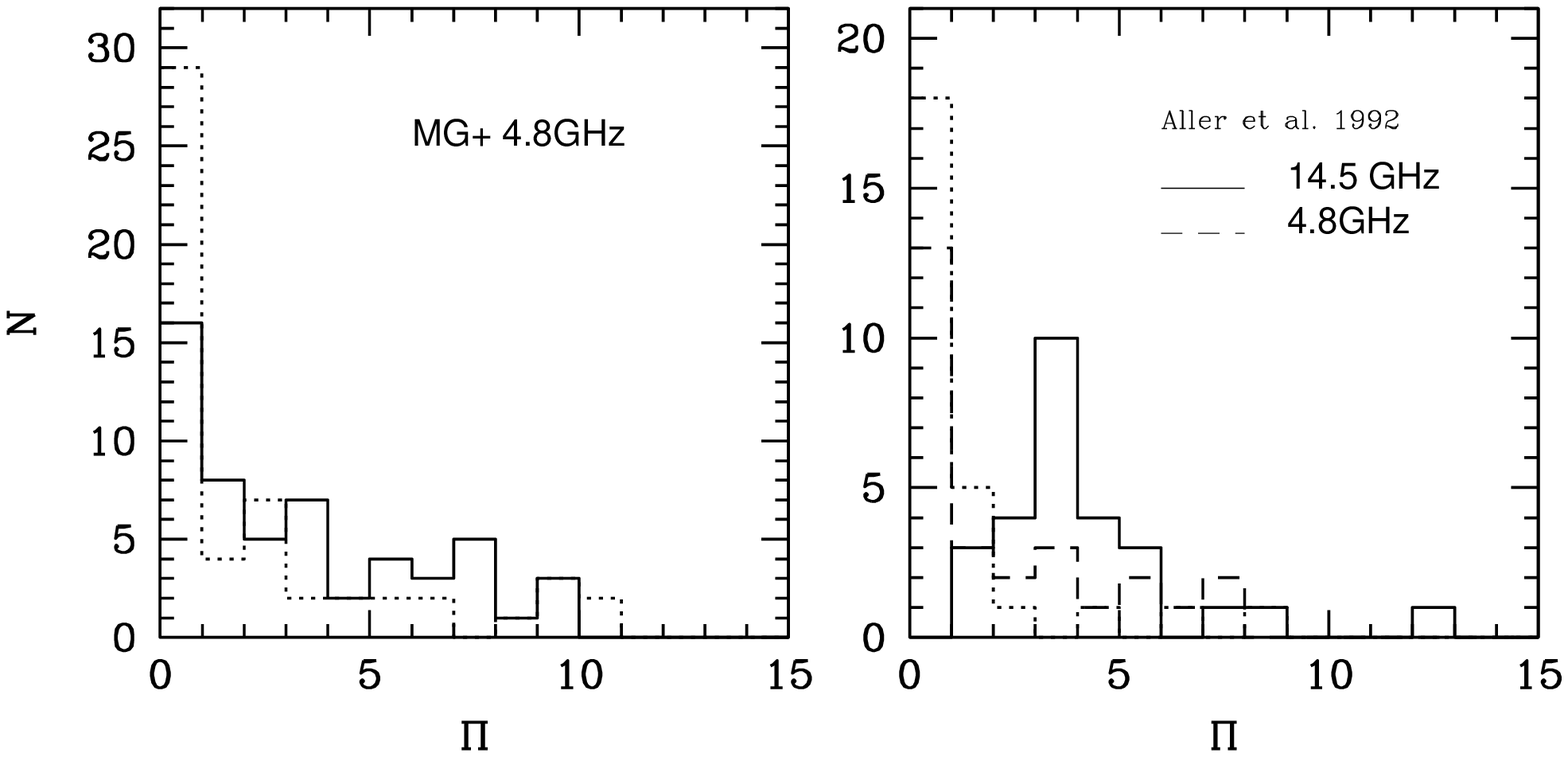}
\caption{Same as in Figure \ref{f6} but for steep--spectrum
sources.} \label{f67}
\end{figure}

\begin{table}
\begin{minipage}{80mm}
\caption{Average polarization degree at various frequencies for
different samples.} \label{t4}
\begin{tabular}{@{}lccccc}
\hline
 & & N & $\nu$(GHz) & $<\Pi>$ & $\sigma$ \\
\hline
 MG+  & flat  & 85 & 1.4 & 2.0 & 1.9 \\
      &       &    & 4.8 & 2.6 & 2.1 \\
      & steep & 54 & 1.4 & 2.4 & 3.1 \\
      &       &    & 4.8 & 3.3 & 2.9 \\
\hline
 \citet{oku93}  & flat  & 89 & 1.4 & 1.8 & 1.7 \\
                &       &    & 10  & 2.7 & 2.1 \\
\hline
 \citet{all92}  & flat  & 35 & 1.4 & 1.8 & 1.7 \\
                &       &    & 4.8 & 2.3 & 2.0 \\
                &       &    & 14.5 & 3.0 & 2.2 \\
                & steep & 27 & 1.4 & 1.3 & 2.1 \\
                &       &    & 4.8 & 2.3 & 2.4 \\
                &       &    & 14.5 & 4.1 & 2.2 \\
\hline
 \citet{lis01}  & flat  & 32 & 1.4 & 2.0 & 1.8 \\
                &       &    & 43  & 3.0 & 2.2 \\
\hline
 \citet{nar98}  & flat  & 26 & 1.4 & 2.4 & 2.0 \\
                &       &    & 273 & 6.6 & 3.0 \\
\hline
VLA Calibrators & flat  & 55 & 1.4 & 1.9 & 2.1 \\
                &       & 48 & 5   & 2.1 & 1.7 \\
                &       & 45 & 8.5 & 2.3 & 2.1 \\
                &       & 47 & 22  & 2.4 & 2.3 \\
                &       & 53 & 43  & 2.8 & 2.4 \\
\hline
\end{tabular}
\medskip
Data at 1.4\,GHz come from the NVSS.
\end{minipage}
\end{table}

Table \ref{t4} compares the values of the mean polarization degree
from samples at different frequencies, while Figures
\ref{f6}--\ref{f67} show the polarization degree distributions for
flat-- and steep--spectrum sources. For flat--spectrum sources,
$<\Pi>$ increases steadily between 1.4 and 14.5\,GHz, although by
only a factor 1.5, and keeps constant between 15 and 43\,GHz in
the samples of \citet{all92} and of \citet{lis01} (note that these
flat--spectrum samples are identical, except for three sources
present at 15\,GHz but not at 43\,GHz). A steady, but weak,
increase of the polarization degree with the frequency is also
found in the case of VLA
calibrators\footnote{http://www.aoc.nrao.edu/$\small{\sim}$smyers/calibration/},
simultaneously observed at 5 frequencies between 5 and 43\,GHz. In
general, data on flat--spectrum sources indicate that this
population is not strongly affected by Faraday depolarization and
that the polarization degree might become frequency independent
already at $\nu\ga15\,$GHz. On the other hand, a significantly
higher mean polarization degree, $<\Pi>=6.6\%$, corresponding to
an increase by nearly a factor of 3 from the value at 1.4\,GHz,
has been found by \citet{nar98} for a sample of blazars at
mm/sub-mm wavelengths, where we are probably observing regions
very close to the active nucleus, whose emission is self--absorbed
at lower frequencies. Thus, apart from the fact that conclusions
based on incomplete samples need always to be taken with care, a
large increase of the polarization degree from 1.4\,GHz to
300\,GHz can be a consequence of the higher uniformity of the
magnetic field in the innermost regions.

Data on steep--spectrum sources are limited to $\nu\la15\,$GHz
(see Table \ref{t1}), and show a clear increase of the
polarization degree with frequency. For the \citet{all92} sample
$<\Pi>=1.3\%$, $\simeq 2.3\%$, $\simeq 4.1\%$ at $\nu=1.4$, 4.8,
14.5\,GHz respectively, while for the MG+ sample $<\Pi>$ changes
from 2.4$\%$ at 1.4\,GHz to 3.3$\%$ at 4.8\,GHz. The factor of 3
increase between 1.4 and 14.5\,GHz is comparable to the result of
M02 from a sample of $\sim 130$ objects, and is compatible with a
strong Faraday depolarization at few--GHz frequencies.

\section{The $\Pi$ distribution of flat--spectrum sources at high
frequencies}
\label{s4}

\begin{figure*}
\includegraphics[width=140mm]{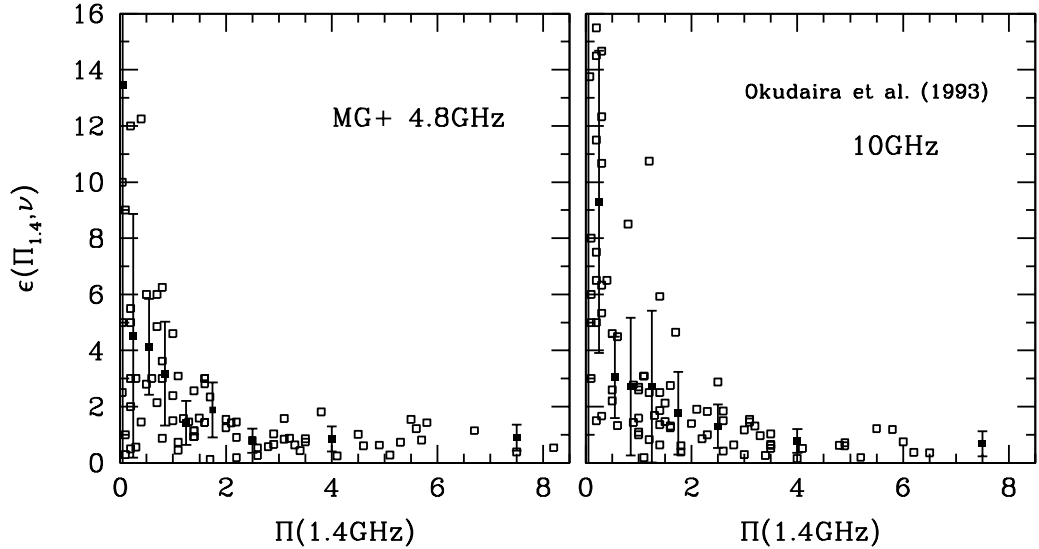}
\caption{Values of $\epsilon$ as a function of $\Pi$ at 1.4\,GHz
for MG+ sources (left plot) and for the \citet{oku93} sample at
10\,GHz (right plot). Points with the error bar give $<\epsilon>$
with its dispersion.} \label{f7}
\end{figure*}

As discussed in Sect. \ref{s2}, the NVSS data are consistent with the
polarization degree of flat--spectrum sources being essentially
independent of flux density. If so, the polarization properties of
bright sources ($S\ga1$\,Jy), for which high frequency data are
available, can be assumed to hold also for much fainter sources.

\begin{table}
\begin{minipage}{80mm}
\caption{Average values of $\epsilon$ and of its variance as a
function of $\Pi_{1.4}$.} \label{t5}
\begin{tabular}{@{}lcccc}
\hline
 & $\Pi_{1.4}$ & N & $<\epsilon>$ & $\sigma$ \\
\hline
 MG+ (4.8\,GHz) & [0,0.1)   & 9  & 13.5 & 13.7 \\
                & [0.1,0.4) & 10 & 4.52 & 4.33 \\
                & [0.4,0.7) & 6  & 4.13 & 1.70 \\
                & [0.7,1)   & 7  & 3.18 & 1.84 \\
                & [1,1.5)   & 11 & 1.42 & 0.79 \\
                & [1.5,2)   & 9  & 1.88 & 0.98 \\
                & [2,3)     & 10 & 0.78 & 0.43 \\
                & [3,5)     & 12 & 0.86 & 0.45 \\
                & $>5$   & 9  & 0.90 & 0.46 \\
\hline
 \citet{oku93} (10\,GHz) & [0,0.1)   & 9  & 27.3 & 31.0 \\
                & [0.1,0.4) & 15 & 9.30 & 5.38 \\
                & [0.4,0.7) & 5  & 3.05 & 1.45 \\
                & [0.7,1)   & 8  & 2.72 & 2.45 \\
                & [1,1.5)   & 14 & 2.72 & 2.70 \\
                & [1.5,2)   & 7  & 1.77 & 1.48 \\
                & [2,3)     & 11 & 1.30 & 0.77 \\
                & [3,5)     & 14 & 0.78 & 0.42 \\
                & $>5$   & 6  & 0.68 & 0.45 \\
\hline
 \citet{all92}  &           &    &      &      \\
  (4.8\,GHz)    & [0,0.1)   & 1  & 1.14 & --   \\
                & [0.1,1)   & 10 & 3.90 & 3.28 \\
                & [1,2)     & 12 & 1.77 & 1.10 \\
                & [2,4)     & 8  & 0.67 & 0.39 \\
                & $>4$   & 4  & 1.08 & 0.47 \\
  (14.5\,GHz)   & [0,0.1)   & 1  & 44.6 & --   \\
                & [0.1,1)   & 10 & 8.94 & 9.38 \\
                & [1,2)     & 12 & 1.98 & 0.84 \\
                & [2,4)     & 8  & 1.46 & 1.37 \\
                & $>4$   & 4  & 0.70 & 0.56 \\
\hline
 \citet{lis01} (43\,GHz) & [0,0.1)   & 0  & --   & --   \\
                & [0.1,1)   & 9  & 7.43 & 4.54 \\
                & [1,2)     & 11 & 2.07 & 1.27 \\
                & [2,4)     & 8  & 1.33 & 0.93 \\
                & $>4$   & 4  & 0.56 & 0.41 \\
\hline
 \citet{nar98}  &           &    &      &      \\
  (273\,GHz)    & [0,0.1)   & 2  & 41.5 & 27.5 \\
                & [0.1,1)   & 6  & 8.10 & 3.48 \\
                & [1,2)     & 6  & 4.96 & 1.95 \\
                & [2,4)     & 6  & 3.05 & 1.28 \\
                & $>4$   & 6  & 1.10 & 0.38 \\
\hline
\end{tabular}
\end{minipage}
\end{table}

First of all, we check whether the polarization of a source at
1.4\,GHz is correlated to its value at higher frequencies. Indeed,
the previously mentioned, frequency dependent effects (Faraday
depolarization and the fact that at different frequencies we may
effectively observe different emitting regions) could spoil such
correlation. If so, low-frequency data would not be representative
of high frequency polarization properties. The Pearson's linear
correlation coefficient $r$ between the polarization at 1.4 and
4.8\,GHz of the sources in the MG+ and \citet{all92} samples is
$\sim0.6$ corresponding to a probability of the null hypothesis,
i.e. no--correlation, of $\la10^{-6}$. The correlation is less
clear in the samples at higher frequencies ($r\sim0.2$ and
null--hypothesis probability of $\ga10^{-2}$). However, these
results are probably affected by the variability of sources,
because we are comparing data taken at different epochs. On the
contrary, a strong correlation ($r>0.5$ and null--hypothesis
probability $<10^{-4}$) between 5 and 43\,GHz is found for the VLA
calibrators, whose polarization is measured simultaneously at all
frequencies.

Consequently, the polarization degree of a source at frequency $\nu$
can be written in term of its value at 1.4\,GHz ($\Pi_{1.4}$) as
\begin{equation}
\Pi(\nu)=\Pi_{1.4}\epsilon(\Pi_{1.4},\nu)\,,
\label{e2}
\end{equation}
where the factor $\epsilon$ is the increase of the polarization
degree from the value at 1.4\,GHz. We find that $\epsilon$ itself
is a function of the polarization degree at 1.4\,GHz:
Fig.~\ref{f7} shows $\epsilon$ as function of $\Pi_{1.4}$ for the
sources in the MG+ and \citet{oku93} samples, while Table \ref{t5}
gives the average $\epsilon$ and its dispersion for different bins
of $\Pi_{1.4}$. For sources with $\Pi_{1.4}$ above a few percent,
$\langle\epsilon\rangle$ is close to unity at all frequencies, but
it can take very large values (up to more than 10) if $\Pi_{1.4}$
is very small.

We can conclude that: $(i)$ the typical intrinsic polarization
degree of flat--spectrum sources is of 2--5$\%$; $(ii)$ high values
of the polarization degree are unlikely (sources with $\Pi\ga10\%$
are rare in the whole range of frequencies analysed), probably due
to the low degree of the uniformity of magnetic fields in radio
sources; $(iii)$ Faraday depolarization is probably the cause of
the large number of sources observed with very low polarization
degree (in fact, flat--spectrum sources with extreme values of RM
have been observed by \citealt{sta98}, \citealt{tay00} and
\citealt{pen00}). Finally, as already pointed out for $<\Pi>$, the
behaviour of $\epsilon(14.5\hbox{GHz})$ and
$\epsilon(43\hbox{GHz})$ is very similar, confirming that the
polarization is weakly frequency dependent for $\nu>15$\,GHz. On
the other hand, $<\epsilon>$ at mm wavelengths is higher than at
GHz frequencies in all $\Pi_{1.4}$ bins.

Using MG+ and \citet{oku93} data, we get a simple analytical
formula for the mean value of $\epsilon(\Pi_{1.4},\nu)$:
\begin{equation}
<\epsilon(\Pi_{1.4},\nu)>=A(\nu)\exp(-3.2\Pi_{1.4}^{0.35})+0.8\,,
\label{e3}
\end{equation}
where the frequency dependence appears only in the coefficient
$A(\nu)$ which takes the values 38, 50 and 108 at $\nu=$4.8, 10
and 273\,GHz respectively. Using these three values, we obtain a
law, $A(\nu)=72\ln(0.75\nu^{0.3}+0.5)$, which allows us to compute
$<\epsilon(\Pi_{1.4},\nu)>$ at every frequency. The large
dispersions around the mean values of $\epsilon$ are described by
a fit similar to Eq. (\ref{e3}), obtained interpolating the values
of the variance of $\epsilon$ in the $\Pi_{1.4}$ bins:
\begin{equation}
\sigma_{<\epsilon>}(\Pi_{1.4},\nu)=B(\nu)\exp(-5\Pi_{1.4}^{0.3})+0.5\,.
\label{e4}
\end{equation}
The factor $B$ is equal to 100 or 210 at $\nu=$4.8 or 10\,GHz,
respectively, and it remains constant at higher frequencies (we
use a linear interpolation at $\nu\le10$\,GHz).

\begin{figure}
\includegraphics[width=84mm]{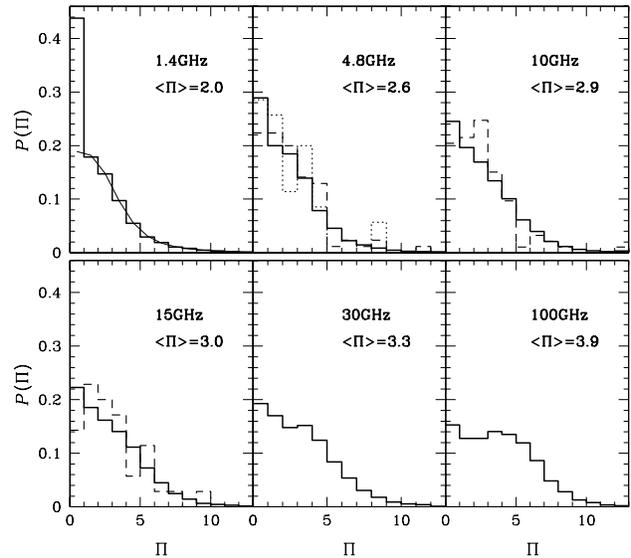}
\caption{Simulated $\Pi$ distributions at several frequencies
between 1.4 and 100\,GHz. The mean value of $\Pi$ is also
indicated. The solid line in the plot at 1.4\,GHz represents the
first term of Eq. (\ref{e1}). We report also the observed $\Pi$
distributions, scaled to the number of simulated sources: 4.8\,GHz
data refer to the MG+ (dashed histogram) and the \citet{all92}
(dotted histogram) samples;  10\,GHz and at 15\,GHz data to the
\citet{oku93} and  \citet{all92} samples respectively.} \label{f9}
\end{figure}

The above expressions for $<\epsilon(\Pi_{1.4},\nu)>$ and for its
variance are used to estimate the polarization degree
distributions of flat--spectrum sources at every frequency in the
range [5,\,300]\,GHz, starting from that of $\Pi_{1.4}$: the
increase factor $\epsilon$ is modelled by a Gaussian distribution
which is cut at negative $\epsilon$ and whose mean value and
variance are given by Eq. (\ref{e3}) and (\ref{e4}). In Figure
\ref{f9} we give examples of the $\Pi$ probability distributions
between 1.4 and 100\,GHz, using a simulated sample of 1000 sources
[the probability distribution at 1.4\,GHz is obtained from Eq.
(\ref{e1})]. The Figure shows also the observed distributions at
$\nu\le15\,$GHz, scaled to the total number of simulated sources;
the good agreement between data and simulations is apparent.
Moreover, the K--S test finds high probabilities ($\ga0.3$) that
the observed distributions are realizations of the probability
distributions resulting from our simulations. A significant agreement
is also found with the recent data at 18\,GHz by \citet{ric03}.

\begin{figure}
\includegraphics[width=75mm]{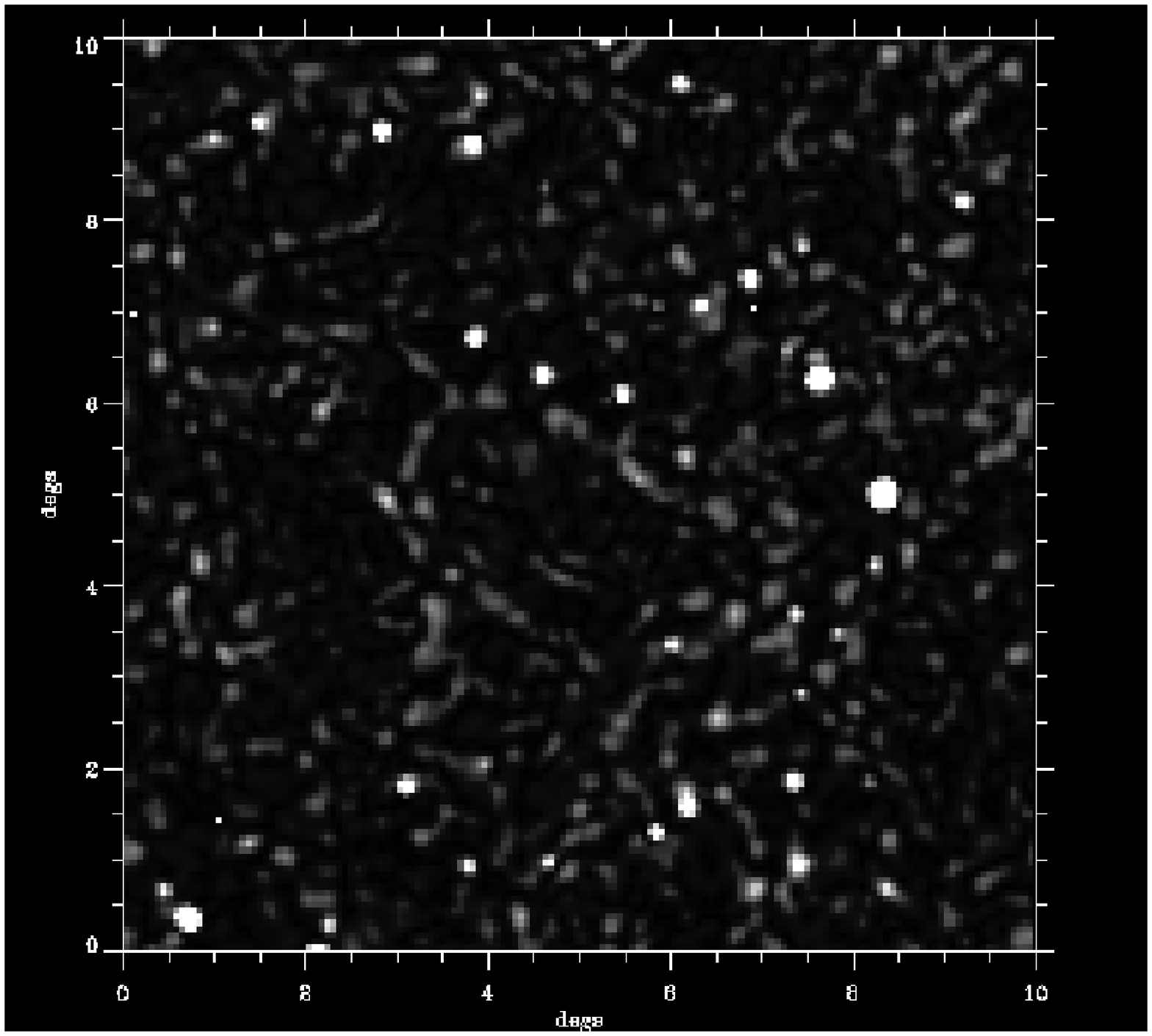}
\includegraphics[width=75mm]{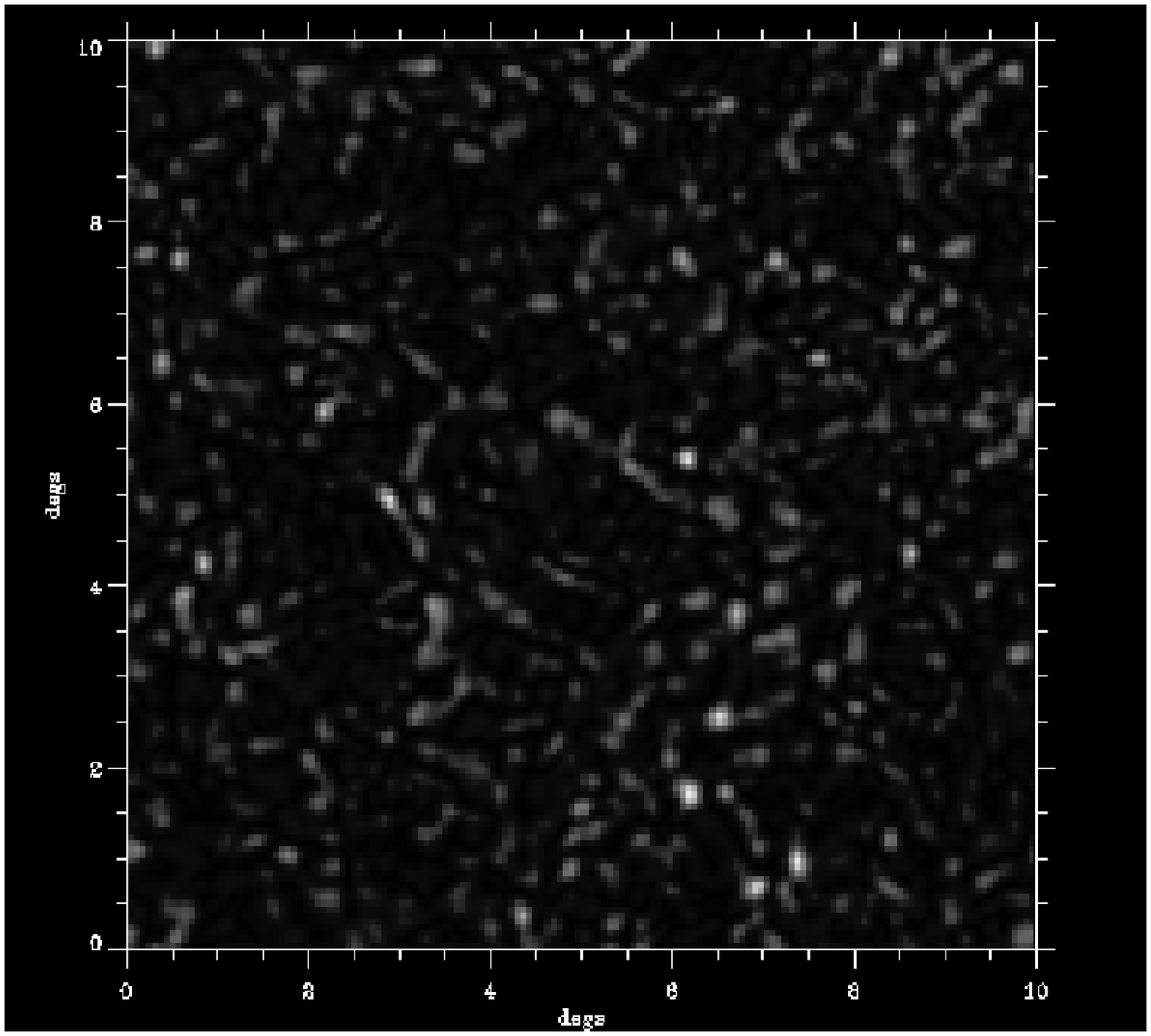}
\caption{Simulated polarized skies at 30 (upper plot) and 100\,GHz
(lower plot) generated including extragalactic radio sources and
CMB radiation (standard CDM model). The sources are far less
conspicuous at 100 GHz than at 30 GHz.} \label{f91}
\end{figure}

Finally, exploiting the T98 model, we generate simulations of the
polarized intensity of the sky at 30 and 100\,GHz including radio
sources and the CMB (see Figure \ref{f91}). We can directly compare
the importance of the contribution of radio sources at the two
frequencies: at 30\,GHz a lot of sources are evident in the map,
superposed to the CMB signal, while at 100\,GHz their presence is far
less conspicuous.

\section{Estimate of Polarization Power Spectra}
\label{s5}

Using the statistical characterization of the polarization degree
described in the previous section, we are able to estimate the
angular power spectrum (APS) of polarization fluctuations due to
extragalactic radio sources. We adopt the T98 model in order to
predict the source number counts at cm and mm wavelengths (the model
number counts are scaled by a factor 0.75 to be in agreement with the
WMAP data).

We assume the spatial distribution of sources to be Poissonian.
The contribution of radio--source clustering to the power spectrum
is, in fact, small and can be neglected if sources are not
subtracted down to very low flux limits (the clustering is noticeable
only at relatively small fluxes, $S\le100$\,mJy; see T98 and
\citealt{tof03}).
Then the power spectrum is independent of the angular scale, and
can be computed in the limit of small scales. Under these
conditions, we have $C_{E\ell}+C_{B\ell}=C_{Q\ell}+C_{U\ell}$
\citep{tuc02}. Moreover, because point sources contribute, on average,
equally to the $E$--, $B$--mode and to $Q$, $U$ power spectra
(\citealt{sel97}, M02), $C_{E\ell} \simeq C_{B\ell} \simeq C_{Q\ell}
\simeq C_{U\ell}$ (in the following we refer to any of these spectra
as the polarization APS).

We estimate the point--source polarization APS in two ways: first
as the product of the temperature power spectrum, $C_{I\ell}$,
and the mean squared polarized intensity; second, directly from
the number counts of $Q$ (or $U$). In both cases, we exploit the
$\Pi$ probability distributions obtained in the previous section.

\citet{teg96} have shown that the APS of intensity fluctuations
produced by Poisson--distributed sources can be written as:
\begin{equation}
C_{I\ell}=N<S^2>=\int_0^{S_c}n(S)S^2\,dS\,,
\label{e6}
\end{equation}
where $N$ and $n(S)$ are, respectively, the total and the
differential number of sources per steradian, and $S_c$ is the
minimum flux density of sources that can be individually detected
and removed.

In analogy to Eq. (\ref{e6}), the $Q$ APS will be
\begin{equation}
C_{Q\ell}=N<Q^2>\,.
\label{e61}
\end{equation}
The Stokes parameter $Q$ can be written in terms of the
polarization degree $p=\Pi/100$ and of the polarization angle in
the chosen reference system, $\phi$, as $Q=Sp\cos(2\phi)$. Because
we find $S$, $p$ and $\phi$ to be independent variables in the
NVSS survey (at least in the case of flat--spectrum sources), we
obtain
\begin{eqnarray}
C_{Q\ell}=N<Q^2> & = & N<S^2p^2\cos^2(2\phi)> \nonumber \\
 & = & N<S^2><p^2><\cos^2(2\phi)> \nonumber \\
 & = & 1/2\,C_{I\ell}<p^2>\,,
\label{e5}
\end{eqnarray}
where the factor $1/2$ arises because of the uniform distribution of
polarization angles.

At a frequency $\nu$, the mean squared value of the polarization
degree, $<p^2_{\nu}>$, is:
\begin{equation}
<p_{\nu}^2>=10^{-4}\int_0^{100}{\mathcal P}(\Pi_{\nu})\Pi_{\nu}^2\,
d\Pi_{\nu},
\label{e7}
\end{equation}
where the probability distribution of $\Pi_{\nu}$ at frequency $\nu$,
${\mathcal P}(\Pi_{\nu})$, is related to ${\mathcal P}(\Pi_{1.4})$
[given in Eq. (\ref{e1})]:
\begin{eqnarray}
{\mathcal P}(\Pi_{\nu}) & = & {\mathcal P}(\Pi_{1.4}){d\Pi_{1.4} \over
d\Pi_{\nu}} \nonumber \\
 & = & {{\mathcal P}(\Pi_{1.4}) \over |\epsilon_{\nu}(\Pi_{1.4})-1.12
[\epsilon_{\nu}(\Pi_{1.4})-0.8]\Pi_{1.4}^{0.35}|}\,.
\label{e8}
\end{eqnarray}
Here we have used the relation
$\Pi_{\nu}=\Pi_{1.4}\epsilon_{\nu}(\Pi_{1.4})$ and Eq. (\ref{e3})
for the mean value of $\epsilon_{\nu}$. For example, we find
$<p^2>^{1/2}=0.041$ at 30\,GHz and $<p^2>^{1/2}=0.046$ at
100\,GHz.

\begin{figure*}
\includegraphics[width=150mm]{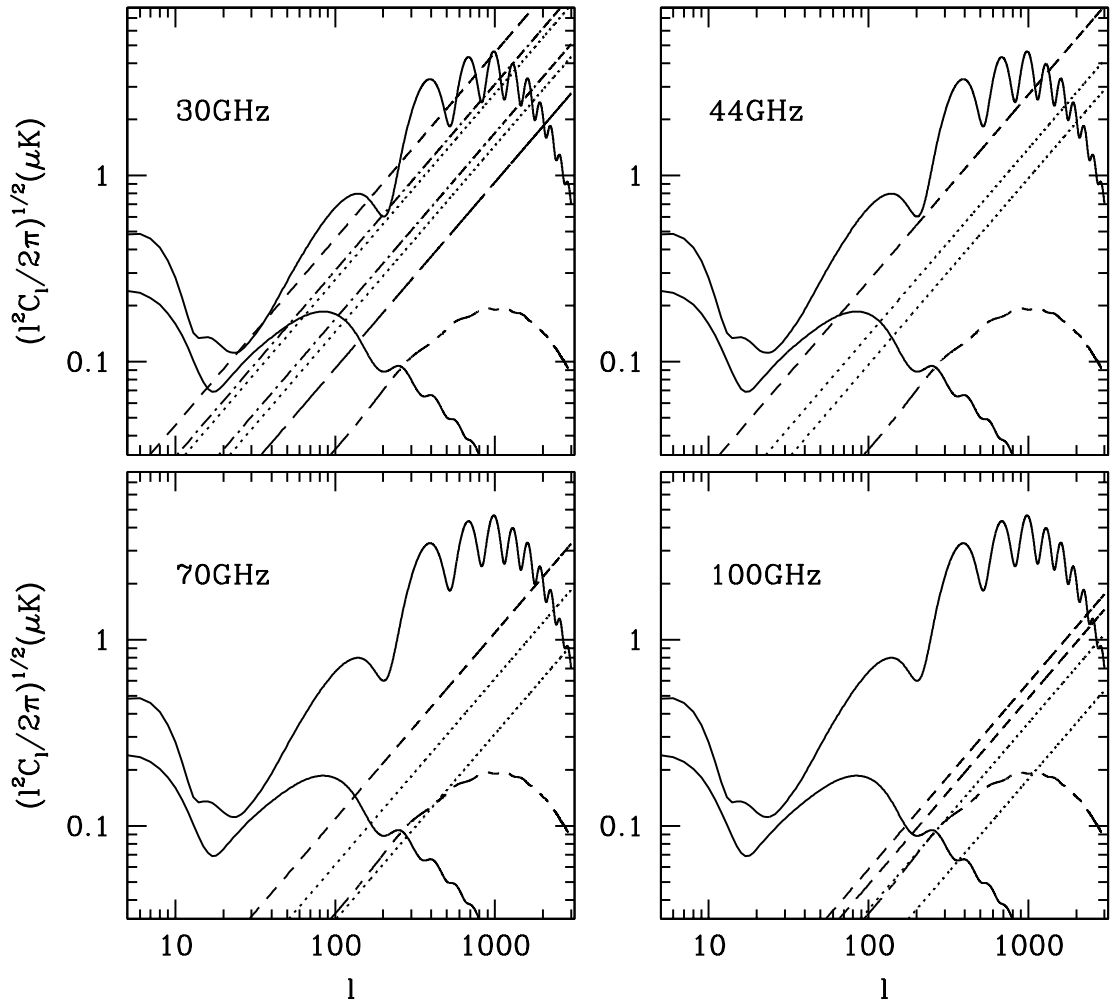}
\caption{Polarization power spectrum produced by extragalactic
radio sources compared to the $E$-- and $B$--mode CMB spectrum
(CDM model with $\Omega_{\Lambda}=0.7$, re-ionization optical
depth $\tau=0.17$ and tensor--to--scalar ratio $r\simeq0.5$). The
short--long dashed lines are the $B$--mode component induced by
gravitational lensing. The CMB and lensing spectra are computed
with CMBFAST \citep{sel96}. The APS for extragalactic radio
sources are computed using the first method with $S_c=1$\,Jy
(upper dotted lines) and $S_c$ estimated by \citet{vie03} for the
Planck mission (200, 400, 250, 200\,mJy at 30, 44, 70, 100\,GHz;
lower dotted lines), and using the second method with
$Q_c=U_c=\sqrt{2}S_c$, where $S_c$ is that from \citet{vie03}
(dashed lines). In the plot at 30\,GHz, the dot--dashed lines
corresponds to the estimates by the first method taking into
account the correction for the contribution of steep--spectrum
sources (see text); the long dashed line is our estimate for the
contribution of undetected sources in the DASI experiment. In the
plot at 100\,GHz, the lower dashed line is the APS computed
assuming the $\Pi$ distribution to become frequency-independent
for $\nu>15\,$GHz.} \label{f10}
\end{figure*}

Finally, it is easy to demonstrate that the cross-correlation $TQ$
spectrum is
\begin{eqnarray}
C_{TQ\ell}=<S^2p\cos(2\phi)>=0.
\label{e81}
\end{eqnarray}
The previous method assumes that sources are removed from
polarization maps using total intensity data. However, not all the
experiments observing the sky polarization are able to measure the
total intensity [see, for example, the experiments: SPOrt
\citep{car03}; COMPASS \citep{far03}; CAPMAP \citep{bar03}]. In
this case, sources have to be detected and removed directly from
$Q$ and $U$  maps. Therefore, we need to compute the $Q$ APS using
the differential sources counts $n(Q)$, as in Eq. (\ref{e6}):
\begin{equation}
C_{Q\ell}=\int_{-Q_c}^{Q_c}n(Q)Q^2\,dQ\,,
\label{e9}
\end{equation}
where $Q_c$ is the lowest value of $Q$ of sources that can be
individually detected and removed from polarization maps. The
counts $n(Q)$ can be obtained from the probability distribution of
the three Stokes parameters that describe the linear polarization,
${\mathcal P}(S,Q,U)$, as
\begin{equation}
n(Q)=\int_{-U_c}^{U_c}dU\,\int_0^{\infty}{\mathcal
P}(S,Q,U)\,dS\,.
\label{e10}
\end{equation}
The probability ${\mathcal P}(S,Q,U)$ is related to the
probability distribution of $S$, of the polarization degree
$p=\sqrt{Q^2+U^2}/S$ and of the polarization angle
$\phi=1/2\arctan(U/Q)$, through
\begin{equation}
{\mathcal P}(S,Q,U)={\mathcal P}(S,p,\phi){\rmn det}{\mathcal J}
\left( \begin{array}{ccc}
S & p & \phi \\
S & Q & U
\end{array} \right)\,,
\label{e11}
\end{equation}
where ${\mathcal J}$ is the Jacobian matrix of the transformation
$(S,Q,U) \rightarrow (S,p,\phi)$:
\begin{equation}
{\rmn det}{\mathcal J}
\left( \begin{array}{ccc}
S & p & \phi \\
S & Q & U
\end{array} \right)={1 \over 2S\sqrt{Q^2+U^2}}\,.
\label{e12}
\end{equation}
As previously noted, the variables $S$, $p$ and $\phi$ are
independent, and therefore
\begin{equation}
{\mathcal P}(S,p,\phi)={\mathcal P}(S){\mathcal P}(p){\mathcal
P}(\phi)=N^{-1}n(S){\mathcal P}(p)/\pi\,;
\label{e12b}
\end{equation}
$n(S)$ is provided by the T98 model and ${\mathcal
P}(p)=100{\mathcal P}(\Pi)$ by Eq. (\ref{e8}). Finally:
\begin{eqnarray}
n(Q) & = & {1 \over \pi}\int_0^{U_c}{dU \over
\sqrt{Q^2+U^2}}\times \nonumber \\
 & \times & \int_0^{\infty}{n(S){\mathcal
P}(\sqrt{Q^2+U^2}/S) \over S}\,dS\,.
\label{e13}
\end{eqnarray}
Figure \ref{f10} shows our estimates of the
polarization APS using the two methods discussed above, at the
frequencies 30, 44, 70 and 100\,GHz.
In each panel, the two dotted curves are computed
using Eq. (\ref{e5}) (i.e., the first method), and $S_c =1\,$Jy
(upper line) or the frequency dependent detection limit (see
caption) obtained by \citet{vie03} for the Planck mission using
the spherical Mexican hat wavelet to remove bright sources (lower
line). The latter limit, multiplied by $\sqrt{2}$, is used for
$Q_c$ (and $U_c$) in Eq. (\ref{e9}) (assuming that the source
removal algorithm has a similar efficiency for polarization maps
as for intensity ones). The factor $\sqrt{2}$ arises if the
noise in temperature and polarization are related by
$\sigma_T^2=\sigma_P^2/2$.
The extrapolation of the polarization degree distribution at 1.4\,GHz
to frequencies between 30 and 100\,GHz is obtained using Eq.
(\ref{e2}) and the mean value of $\epsilon$, Eq. (\ref{e3}).
The uncertainty on the polarization APS due to the high--frequency
extrapolation can be estimated by different realizations of
$\Pi$ distributions generated following the recipe we describe in
section \ref{s4}. We find that the variance is small, a few percent of
the average spectrum (the number of sources is large enough to assure
the sample variance to be negligible).

As discussed in section \ref{s4}, if the data of \citet{nar98} are
disregarded, the polarization degree appears to be weakly dependent on
the frequency for $\nu>15$\,GHz. Assuming the $\Pi$ distribution to be
constant above 15\,GHz, the amplitude of the polarization APS
decreases weakly with respect to previous estimates: $\sim10\%$ at
30\,GHz and $\sim30\%$ at 100\,GHz (the differences are visibly
appreciable only in the 100\,GHz panel of Fig. \ref{f10}). These
latter estimates can be considered as lower limits and, together with
the ones using the high--frequency extrapolation, provide us with two
boundaries within which we expect to find the real polarization APS.

We have verified that the results from the two methods agree when
the bright source removal is equivalent.  Otherwise, as
illustrated by Fig. \ref{f10}, estimates based on Eq. (\ref{e5})
are significantly lower than those using Eq. (\ref{e9}): the
difference is nearly a factor 10 if we take $Q_c=U_c=\sqrt{2}S_c$
(in the real case, however, we expect that $Q_c=U_c>\sqrt{2}S_c$,
further increasing the difference). This is easily understood:
adopting the limit on total flux density we are subtracting many
more sources than we can do using only polarization maps.

The previous estimates are obtained assuming that the polarization
degree distribution for steep--spectrum sources is the same as for
the flat--spectrum ones, although the results reported in Table
\ref{t4} and by M02 indicate that steep--spectrum sources are, on
average, more polarized, at least at low frequencies. However, the
contribution of steep--spectrum sources to the polarization APS is
anyway very small at $\nu>40\,$GHz.

The likely correction to our results at 30 GHz can be estimated as
follows. The VSA survey at 15\,GHz \citep{wal03} finds a 25 percent of
steep--spectrum sources for $S_{15}\ge100\,$mJy and a 44 percent for
$S_{15}\ge25\,$mJy, corresponding to a contribution to $C_{I\ell}$
of 30$\%$ if $S_c=1\,$Jy and of 43$\%$ if $S_c=200\,$mJy (we
assume that the fraction of steep--spectrum sources increases to
75$\%$ if $S_{15}<25\,$mJy). Their contribution to the APS is
reduced by a factor 3 at 30\,GHz if their average spectral index
is $-0.8$ between 15 and 30\,GHz (a rather conservative assumption
since the high frequency spectra steepen due to electron ageing
effects). From the NVSS catalogue we find that
$<\Pi_{1.4}^2>^{1/2}\simeq2.6\%$ for steep--spectrum sources with
$S_{1.4}\ge200\,$mJy. Now, assuming that the polarization degree
increases, on average, by a factor of 3 from 1.4\,GHz to high
frequencies, we obtain that our estimates of the polarization APS
by Eq. (\ref{e5}) must be multiplied at 30\,GHz by 1.3 or 1.4 in
the case that $S_c=1\,$Jy or 200\,mJy respectively (dot--dashed
lines in the 30\,GHz plot of Figure \ref{f10}).

\section{Discussions and conclusions}
\label{s6}

The main goal of the present paper is to provide estimates of the
contamination of CMB polarization maps by extragalactic radio
sources. This task is particularly complicated because of the lack
of polarimetric data at the frequencies where the CMB is observed.

The richest data on the polarization properties of radio sources
comes from the NVSS survey, which provides a very large and
complete catalogue of extragalactic sources at 1.4\,GHz.
Exploiting the spectral information obtained combining the NVSS
with the GB6 survey at 4.8\,GHz, we have analyzed the properties
of steep-- and flat--spectrum sources. The former population shows
an anti--correlation of the polarization degree with the flux
density, already pointed out by M02. On the contrary, for
flat--spectrum sources the distributions of the polarization
degree for different flux-density ranges  do not show any
significant variation and its mean value is constant and quite
low, around $2\%$. For this class of sources we find a fit that
accurately describes  the polarization degree distribution for all
sources with $S\ge100$\,mJy.

Exploiting the available data at $\nu>1.4$\,GHz, we have
investigated how the polarization degree of sources varies with
frequency. Data on steep--spectrum sources, available only up to
15\,GHz, highlight a strong increase of the mean polarization
degree with frequency, consistent with substantial Faraday
rotation measures. On the other hand, the mean polarization degree
of flat--spectrum sources shows only a weak increase between 1.4
and 15\,GHz, and appears to remain essentially constant at higher
frequencies. However, the high values of the polarization degree
found at mm wavelengths by \citet{nar98} may be an indication that
new, more polarized, components show up there.

For flat--spectrum sources, we have derived an analytic function
allowing us to extrapolate to any frequency the distribution of
polarization degrees at 1.4\,GHz. We notice that only sources with
$\Pi_{1.4}<1\%$ increase significantly their polarization degree
between 5 and 43\,GHz. These sources may be the only ones with
really strong Faraday depolarization at 1.4\,GHz. In general, at
$\nu\ga5\,$GHz the polarization degree of flat--spectrum sources
is typically of a few percent (it rarely exceeds $10\%$), and is
weakly dependent on the frequency.

Using this analytic function, we estimate the power spectrum of
polarization fluctuations induced by extragalactic radio sources,
by means of the two methods described in Section \ref{s5}. In
particular, we consider for the first time the case whereby the
removal of the brightest sources can be only done using
polarization data. In this case, the source subtraction is far
less efficient than if total flux data can be used and, as a
consequence, the amplitude of the source power spectrum can be up
to a factor of 10 higher.  Total intensity data are, therefore,
essential information in order to reduce the contamination of CMB
polarization maps. In Figure \ref{f10} we show the estimated
contribution of undetected radio sources to the polarization
signal observed by the DASI experiment at 30\,GHz: if indeed all
the sources with the flux density higher than 50\,mJy have been
removed, the residual contamination is extremely small.

We have focused our analysis on flat--spectrum sources, since they
are the dominant population at least up to 100\,GHz.
Steep--spectrum sources are taken into account only at 30\,GHz,
while they are disregarded at higher frequencies. About the GPS
sources, large uncertainties are still present on their abundance.
The WMAP data at 33\,GHz detected 16 sources with inverted spectra
($\alpha<-0.4$) over a total of 155 sources with $S>1.2$\,Jy. 
The multifrequency data summarized by \citet{tru03}
indicate that most of them are likely to be
blazars (flat-spectrum quasars or BL Lacs) whose radio emission is
dominated by a single emitting region (a knot in the jet) caught
during a flare. Thus, the surface density of bright GPS sources
peaking at high radio frequencies is likely to be several times
lower than predicted by the \citet{dez00} models, even in
the case of an intrinsic distribution of peak frequencies not
extending above $\sim 200\,$GHz. According to the latter model
(which, in the light of WMAP sources, provides rather generous
upper limits), at $\nu = 100\,$GHz, the contribution of GPS
sources to the total intensity APS is at the few percent level.
Furthermore, the polarization degree of bona-fide GPS sources is
very low at cm wavelengths. GPS galaxies generally have fractional
polarization below 0.3\% (many are undetected below 0.1\% levels);
GPS quasars have higher polarization than galaxies, but lower (at
6 cm) than non-GPS quasars (\citealt{ode90}; \citealt{sta98};
\citealt{sta99}). Thus, the
contribution of GPS sources to the polarization APS is likely to
be even lower than to the total intensity APS.

It is interesting to compare our results on extragalactic radio
sources with the E-- and B--mode spectra of the CMB radiation (see
Figure \ref{f10}). First of all, we notice that the CMB E--mode
APS increases with $\ell$ in a similar way to the point--source
spectrum in the range $30\la\ell\la10^3$. At these angular scales,
radio sources yield a relevant contribution to the E-mode
polarization only for $\nu\la40$\,GHz; in any case, also at these
frequencies, they are not a serious problem for measurements of
the CMB E--mode APS when the brightest sources are removed. At
higher frequencies the cosmological signal is dominant at least up
to $\ell\sim2000$ or more.

On the other hand, extragalactic radio sources can be a critical
factor for the detection of the CMB B--mode component: the
cosmological $C_{B\ell}$ is characterized by a peak at
$\ell\sim100$ whose amplitude is directly related to the inflation
model and, in particular, to the ratio between the amplitude of
tensor and scalar perturbations. It is well known that an
important limitation to the detectability of such peak comes from
B--modes induced by gravitational lensing \citep{kno02},
especially on scales $\ell\ga100$. However, our analysis
highlights that, at least for $\nu\la100\,$GHz, B--modes produced
by extragalactic radio sources are even more critical. A better
understanding of polarization properties of dust in the Milky Way
and in external galaxies is necessary to establish whether
frequencies higher than 100\,GHz may be more suitable to
investigate the B--mode polarization.

\vskip 0.7truecm \noindent {\it Acknowledgements}. We gratefully
acknowledge the financial support provided through the European
Community's Human Potential Programme under contract
HPRN-CT-2000-00124, CMBNET, and through the Spanish Ministerio de
Ciencia y Tecnologia, reference ESP2002-04141-C03-01. GDZ has been
partially supported by ASI and MIUR (COFIN 2002).


\begin{thebibliography}{}

\bibitem[\protect\citeauthoryear{Aller, Aller \& Hughes}{Aller et
al.}{1992}]{all92}
Aller M. F., Aller H. D., Hughes P. A., 1992, ApJ, 399, 16.
\bibitem[\protect\citeauthoryear{Aller et al.}{1999}]{all99}
Aller M. F., Aller H. D., Hughes P. A., Latimer G. E., 1999, ApJ, 512,
601.
\bibitem[\protect\citeauthoryear{Baccigalupi et al.}{2001}]{bac01}
Baccigalupi C., Burigana C., Perrotta F., De Zotti G., La Porta
L., Maino D., Maris M., Paladini R., 2001, A\&A,  372, 8.
\bibitem[\protect\citeauthoryear{Barkats}{2003}]{bar03}
Barkats D., 2003, in {\it The Cosmic Microwave Background and
its Polarization}, Hanany S., Olive K.A. (eds.), New Astronomy
Reviews, 47, 1077.
\bibitem[\protect\citeauthoryear{Bennett et al.}{1986}]{ben86}
Bennett C. L., Lawrence C. R., Burke B. F., Hewitt J. N., Mahoney J.,
1986, ApJS, 61, 1.
\bibitem[\protect\citeauthoryear{Bennett et al.}{2003a}]{ben03a}
Bennett C. L., et al., 2003a, ApJ, 583, 1.
\bibitem[\protect\citeauthoryear{Bennett et al.}{2003b}]{ben03b}
Bennett C. L., et al., 2003b, ApJS, 148, 97.
\bibitem[\protect\citeauthoryear{Beno{\^{\i}}t et al.}{2003}]{ben03}
Beno{\^{\i}}t A., et al., 2003, A\&A, submitted, astro-ph/0306222.
\bibitem[\protect\citeauthoryear{Bruscoli et al.}{2002}]{bru02}
Bruscoli M., Tucci M., Natale V., Carretti E., Fabbri R., Sbarra C.,
Cortiglioni S., 2002, NewA, 7, 171.
\bibitem[\protect\citeauthoryear{Carretti et al.}{2003}]{car03}
Carretti E., et al., 2003, in {\it Polarimetry in Astronomy}, Fineschi
S. (eds.), SPIE Symposium on Astronomical Telescopes and
Instrumentation, Waikoloa, Hawaii, SPIE Conf. Proc. 4843, 305;
astro--ph/0212067.
\bibitem[\protect\citeauthoryear{Cecchini et al.}{2002}]{cec02}
Cecchini S., Cortiglioni S., Sault R., Sbarra C. (eds.), 2002,
{\it Astrophysical Polarized Backgrounds}, AIP Conf. Proc. 609.
\bibitem[\protect\citeauthoryear{Condon et al.}{1998}]{con98}
Condon J. J., Cotton W. D., Greisen E. W., Yin Q. F., Perley R. A.,
Taylor G. B., Broderick J. J., 1998, AJ, 115, 1693.
\bibitem[\protect\citeauthoryear{De Zotti et al.}{1999}]{dez99}
De Zotti G., Gruppioni C., Ciliegi P., Burigana C., Danese L., 1999,
NewA, 4, 481.
\bibitem[\protect\citeauthoryear{De Zotti et al.}{2000}]{dez00}
De Zotti G., Granato G.L., Silva L., Maino D., Danese L., 2000, A\&A,
354, 467.
\bibitem[\protect\citeauthoryear{Eichendorf \& Reinhardt}{1979}]{eic79}
Eichendorf W., Reinhardt M., 1979, Ap\&SS, 61, 153.
\bibitem[\protect\citeauthoryear{Farese et al.}{2003}]{far03}
Farese P.C., et al., 2003, in {\it The Cosmic Microwave Background and
its Polarization}, Hanany S., Olive K.A. (eds.), New Astronomy
Reviews, 47, 1033.
\bibitem[\protect\citeauthoryear{Fosalba et al.}{2002}]{fos02}
Fosalba P., Lazarian A., Prunet S., Tauber J. A., 2002, ApJ, 564,
762.
\bibitem[\protect\citeauthoryear{Gregory et al.}{1996}]{gre96}
Gregory P. C., Scott W. K., Douglas K., Condon J. J., 1996, ApJS, 103,
427.
\bibitem[\protect\citeauthoryear{Griffith et al.}{1990}]{gri90}
Griffith M., Langston G., Heflin M., Conner S., Lehar J., Burke B.,
1990, ApJS, 74, 129.
\bibitem[\protect\citeauthoryear{Griffith et al.}{1991}]{gri91}
Griffith M., Heflin M., Conner S., Burke B., Langston G., 1991, ApJS,
75, 801.
\bibitem[\protect\citeauthoryear{Hanany \& Olive}{2003}]{han03}
Hanany S., Olive K.A. (eds.), 2003, {\it The Cosmic Microwave
Background and its Polarization}, New Astronomy Reviews.
\bibitem[\protect\citeauthoryear{Knox \& Song}{2002}]{kno02}
Knox L., Song Y.-S., 2002, Phys. Rev. Lett., 89, 011303.
\bibitem[\protect\citeauthoryear{Kovac et al.}{2002}]{kov02}
Kovac J. M., Leitch E. M., Pryke C., Carlstrom J. E., Halverson N. W.,
Holzapfel W. L., 2002, Nature, 420, 772.
\bibitem[\protect\citeauthoryear{Kogut et al.}{2003}]{kog03}
Kogut A., et al., 2003, ApJS, 148, 161.
\bibitem[\protect\citeauthoryear{Kamionkowski, Kosowsky \&
Stebbins}{Kamionkowski et al.}{1997}]{kam97}
Kamionkowski M., Kosowsky A., Stebbins A., 1997, Phys. Rev. D, 55,
7368.
\bibitem[\protect\citeauthoryear{Langston et al.}{1990}]{lan90}
Langston G. I., Heflin M. B., Conner S. R., Lehar J., Carrilli C. L.,
Burke B. F., 1990, ApJS, 72, 621.
\bibitem[\protect\citeauthoryear{Lister}{2001}]{lis01}
Lister M. L., 2001, A\&AS, 198, 7908.
\bibitem[\protect\citeauthoryear{Mason et al.}{2003}]{mas02}
Mason B.S., et al., 2003, ApJ, 591, 540.
\bibitem[\protect\citeauthoryear{Mesa et al.}{2002}]{mes02}
Mesa D., Baccigalupi C., De Zotti G., Gregorini L., Mack K.-H.,
Vigotti M., Klein U., 2002, A\&A, 396, 463 (M02).
\bibitem[\protect\citeauthoryear{Nartallo et al.}{1998}]{nar98}
Nartallo R., Gear W. K., Murray A. G., Robson E. I., Hough J. H.,
1998, MNRAS, 297, 667.
\bibitem[\protect\citeauthoryear{O'Dea}{1989}]{ode89}
O'Dea C.P., 1989, A\&A, 210, 35.
\bibitem[\protect\citeauthoryear{O'Dea et al.}{1990}]{ode90}
O'Dea C.P., Baum S.A., Stanghellini C., Morris G.B., Patnaik A.R.,
Gopal--Krishna, 1990, A\&ASS, 84, 549.
\bibitem[\protect\citeauthoryear{Okudaira et al.}{1993}]{oku93}
Okudaira A., Tabara H., Kato T., Inoue M., 1993, PASJ, 45, 153.
\bibitem[\protect\citeauthoryear{Pentericci et al.}{2000}]{pen00}
Pentericci L., Van Reeven W., Carilli C.L., R$\ddot{o}$tgering H.J.A.,
Miley G.K., 2000, A\&AS, 145, 121.
\bibitem[\protect\citeauthoryear{Perley}{1982}]{per82}
Perley R. A., 1982, AJ, 87, 6.
\bibitem[\protect\citeauthoryear{Pearson \& Readhead}{1981}]{pea81}
Pearson T.J., Readhead A.C.S., 1981, ApJ,  248, 61
\bibitem[\protect\citeauthoryear{Pearson \& Readhead}{1988}]{pea88}
Pearson T.J., Readhead A.C.S., 1988, ApJ,  328, 114
\bibitem[\protect\citeauthoryear{Prunet et al.}{1998}]{pru98}
Prunet S., Sethi S. K., Bouchet F. R., Miville--Deschenes M.A., 1998,
A\&A, 339, 187.
\bibitem[\protect\citeauthoryear{Ricci et al.}{2003}]{ric03}
Ricci R., Prandoni I., Gruppioni C., Sault R.J., De Zotti G., 2003,
A\&A, accepted; astro--ph/0312163.
\bibitem[\protect\citeauthoryear{Rudnick et al.}{1985}]{rud85}
Rudnick L., Jones T. W., Aller H. D., Aller M. F., Hodge P. E., Owen
F. N., Fiedler R. L., Puschell J. J., Bignell R. C., 1985, ApJS, 57,
693.
\bibitem[\protect\citeauthoryear{Seljak}{1997}]{sel97}
Seljak U., 1997, ApJ, 482, 6.
\bibitem[\protect\citeauthoryear{Seljak \& Zaldarriaga}{1996}]{sel96}
Seljak U., Zaldarriaga M., 1996, ApJ, 469, 437.
\bibitem[\protect\citeauthoryear{Simard--Normandin, Kronberg \&
Button}{Simard--Normandin et al.}{1981}]{sim81}
Simard--Normandin M., Kronberg P. P., Button S., 1981, ApJS, 45, 97.
\bibitem[\protect\citeauthoryear{Stanghellini}{1999}]{sta99}
Stanghellini, C., 1999, Mem. SAIt, 70, 117. 
\bibitem[\protect\citeauthoryear{Stanghellini et al.}{1998}]{sta98}
Stanghellini C., O'Dea C. P., Dallacasa D., Baum S. A., Fanti R.,
Fanti C., A\&AS, 1998, 131, 303.
\bibitem[\protect\citeauthoryear{Tabara \& Inoue}{1980}]{tab80}
Tabara H., Inoue M., 1980, A\&AS, 39, 379.
\bibitem[\protect\citeauthoryear{Taylor}{2000}]{tay00}
Taylor G. B., 2000, ApJ, 533, 95.
\bibitem[\protect\citeauthoryear{Taylor et al.}{2001}]{tay01}
Taylor A.C., Grainge K., Jones M. E., Pooley G. G., Saunders
R. D. E., Waldram E. M., 2001, MNRAS, 327, L1.
\bibitem[\protect\citeauthoryear{Tegmark \& Efstathiou}{1996}]{teg96}
Tegmark M., Efstathiou G., 1996, MNRAS, 281, 1297.
\bibitem[\protect\citeauthoryear{Toffolatti et al.}{1998}]{tof98}
Toffolatti L., Arg{\"u}eso--Gomez F., De Zotti G., Mazzei P.,
Franceschini L., Danese L., Burigana C., 1998, MNRAS, 297,117 (T98).
\bibitem[\protect\citeauthoryear{Toffolatti et al.}{2003}]{tof03}
Toffolatti L., et al., 2003, in preparation.
\bibitem[\protect\citeauthoryear{Trushkin}{2003}]{tru03}
Trushkin, S.A., 2003, Bull. Spec. Astrophys. Obs. N. Caucasus, 55,
90; astro-ph/0307205.
\bibitem[\protect\citeauthoryear{Tucci et al.}{2000}]{tuc00}
Tucci M., Carretti E., Cecchini S., Fabbri R., Orsini M., Pierpaoli E.,
2000, NewA, 5, 181.
\bibitem[\protect\citeauthoryear{Tucci et al.}{2002}]{tuc02}
Tucci M., Carretti E., Cecchini S., Nicastro L., Fabbri R., Gaensler
B.M., Dickey J.M., McClure--Griffiths N.M., 2002, ApJ, 579, 607.
\bibitem[\protect\citeauthoryear{Vielva et al.}{2003}]{vie03}
Vielva P., Mart\'{\i}nez--Gonz\'alez E., Gallegos J.E., Toffolatti L.,
Sanz J.L., 2003, MNRAS, 344, 89.
\bibitem[\protect\citeauthoryear{Waldram et al.}{2003}]{wal03}
Waldram E.M., Pooley G.G., Grainge K.J.B., Jones M.E., Saunders
R.D.E., Scott P.F., Taylor A.C., 2003, MNRAS, 342, 915.
\bibitem[\protect\citeauthoryear{Zaldarriaga \& Seljak}{1997}]{zal97}
Zaldarriaga M., Seljak U., 1997, Phys. Rev. D, 55, 1830.
\bibitem[\protect\citeauthoryear{Zaldarriaga \& Seljak}{1998}]{zal98}
Zaldarriaga M., Seljak U., 1998, Phys. Rev. D, 58, 3003.
\bibitem[\protect\citeauthoryear{Zukowski et al.}{1999}]{zuk99}
Zukowski E. L., Kronberg P. P., Forkert T., Wielebinski R., 1999,
A\&AS, 135, 571.
\end{thebibliography}
\end{document}